\documentclass[twocolumn]{aastex63}

\usepackage{graphicx}
\usepackage{graphics}
\usepackage{amssymb}
\usepackage{bm}
\usepackage{times}
\usepackage[flushleft]{threeparttable}
\usepackage{booktabs}
\usepackage{footnote}
\usepackage{multirow}

\newcommand{\kms}{km~s$^{-1}$}

\newcommand{\ha}{\ensuremath{{\rm H}\alpha}}

\newcommand{\arcs}{\ensuremath{^{\prime\prime}}}

\newcommand{\hicm}{H{\sc i} 21 cm}
\newcommand{\hi}{H{\sc i}}
\definecolor{cerulean}{rgb}{0.0, 0.48, 0.65}
\definecolor{red}{rgb}{1.0, 0.0, 0.0}

\definecolor{green}{rgb}{0.0, 0.6, 0.6}

\shorttitle{Unusual gas structure in  GRB 171205A host}
\shortauthors{Arabsalmani et al.}
\begin{document}


\title
{Unusual  gas structure in an otherwise normal spiral galaxy hosting GRB 171205A / SN 2017iuk}

\correspondingauthor{Maryam Arabsalmani, Vera Rubin Fellow}
\email{maryam.arabsalmani@origins-cluster.de}

\author{M. Arabsalmani}, 
\affiliation{Excellence Cluster ORIGINS, Boltzmannstra{\ss}e 2, 85748 Garching, Germany}
\affiliation{Ludwig-Maximilians-Universit\"at, Schellingstra{\ss}e 4, 80799 M\"unchen, Germany}

\author{S. Roychowdhury}
\affiliation{University Observatory Munich (USM), Scheinerstra{\ss}e 1, 81679 M\"unchen, Germany}

\author{F. Renaud}
\affiliation{Department of Astronomy and Theoretical Physics, Lund Observatory, Box 43, 221 00 Lund, Sweden}

\author{A. Burkert}
\affiliation{University Observatory Munich (USM), Scheinerstra{\ss}e 1, 81679 M\"unchen, Germany}
\affiliation{Max-Planck-Institut f\"ur extraterrestrische Physik (MPE), Giessenbachstr. 1, 85748 Garching, Germany}

\author{E. Emsellem}
\affiliation{European Southern Observatory, Karl-Schwarzschild-Stra{\ss}e 2, 85748 Garching, Germany}
\affiliation{Univ. Lyon, Univ. Lyon1, ENS de Lyon, CNRS, Centre de Recherche Astrophysique de Lyon, UMR5574, 69230 Saint-Genis-Laval, France}

\author{E. Le Floc'h}
\affiliation{IRFU, CEA, Universit\'e Paris-Saclay, F-91191 Gif-sur-Yvette, France}
\affiliation{Universit\'e Paris Diderot, AIM, Sorbonne Paris Cit\'e, CEA, CNRS, F-91191 Gif-sur-Yvette, France}

\author{E. Pian}
\affiliation{INAF, Astrophysics and Space Science Observatory, via P. Gobetti 101, 40129 Bologna, Italy}


\begin{abstract}
We study the  structure of atomic hydrogen (\hi) in the  host galaxy of GRB 171205A / SN 2017iuk at $z=0.037$ through \hicm\ emission line observations with the Karl G. Jansky Very Large Array. These observations reveal  unusual morphology and kinematics of the \hi\ in this otherwise apparently normal galaxy. High column density, cold \hi\ is absent from an extended North-South region passing by the optical centre of the galaxy, but instead is extended towards the South, on both sides of the galaxy.  
Moreover, the  \hi\  kinematics  
do not show a continuous change along the major axis of the galaxy as expected in a classical rotating disk. 
We explore several scenarios to explain the \hi\ structure and kinematics in the galaxy:   feedback from a central starburst and/or an active galactic nucleus, ram pressure stripping, accretion, and tidal interaction from a companion galaxy. All of these options are ruled out. 
The most viable remaining explanation    is the  penetrating passage of a  satellite through the disk  only a few Myr ago, redistributing   the \hi\  in the GRB host without yet affecting its stellar distribution. 
It can also lead to  the rapid formation of peculiar stars due to a violent induced  shock. 
The location of GRB 171205A  in the vicinity of the distorted area suggests that its  progenitor star(s)  originated in 
extreme conditions that share  the same origin as the peculiarities in \hi. This could  explain the  atypical location of GRB 171205A in its host galaxy. 

\end{abstract}




\section{Introduction} 
\label{sec:int}

The large-scale dynamics of star-forming galaxies can be traced through the distribution and kinematics of their atomic hydrogen (\hi) reservoirs.
This is particularly so because the \hi\ reservoir of star-forming (both spiral and dwarf) galaxies typically extend much beyond their stellar  disks \citep[][]{Begum08-2008MNRAS.386.1667B, Walter08-2008AJ....136.2563W}. This extended \hi\ helps register signatures of varied dynamical mechanisms  like tidal interactions and mergers \citep[e.g.][]{Sancisi99-1999Ap&SS.269...59S}, gas infall \citep[e.g.][]{Sancisi08-2008A&ARv..15..189S}, and various mechanisms that can affect the content, distribution, and kinematics of gas in galaxies in group and cluster  environments \citep[e.g. see][for a detailed discussion on such mechanisms]{Chung09-2009AJ....138.1741C}.

Recent studies indicate  a possible link between rare galaxy dynamics and the extreme conditions that ignite  the bright explosions of massive stars:   long-duration Gamma Ray Bursts (GRBs), commonly associated with energetic supernovae (SNe),  and superluminous supernovae \citep[SLSNe;][]{Arabsalmani19-2019MNRAS.485.5411A, Arabsalmani19-2019ApJ...882...31A, Roychowdhury19-2019MNRAS.485L..93R}. These energetic events are believed to  occur at the end of the life-cycle of extremely massive stars ($\rm M \gtrsim 40-100 M_{\odot}$) and are therefore quite rare \citep[see][and references therein]{Woosley06-2006ApJ...637..914W, Gal-Yam12-2012Sci...337..927G}. The study of \hi\ structure in the closest known GRB host (GRB 980425 / SN 1998bw, $z=0.0087$)  revealed it to be a collisional ring galaxy, with the GRB residing within the ring 
\citep[][]{Arabsalmani15-2015MNRAS.454L..51A, Arabsalmani19-2019MNRAS.485.5411A}.   Such rings are ideal places for formation of massive star clusters in which very massive stars could form. The resolved molecular gas studies of the host  galaxy showed the presence of local starburst modes of star formation \citep[][]{Arabsalmani20-2020ApJ...899..165A}, again ideal for formation of massive star clusters. \citet[][]{Roychowdhury19-2019MNRAS.485L..93R} found the  mysterious   transient AT2018cow ($z=0.0141$) to be also within a  broken-ring of \hi\ in its host galaxy.
Furthermore,   \citet[][]{Arabsalmani19-2019ApJ...882...31A} found the location of SLSN PTF10tpz ($z=0.0399$) to be within very dense gas clouds, formed by the interaction of gas flows and due to the dynamics of the bar in the host galaxy. 
Such  rare  dynamics (due to external/internal effects) could be the  factors singling  out  the host galaxies of these bright transients,  and  result in the  formation of their    massive progenitor stars. In other words, these  rare and energetic transients could be  the tracers of  gas-rich  galaxies with extreme dynamics.

In this paper we  present a detailed study of  \hi\ in the host galaxy of GRB 171205A / SN 2017iuk at $z=0.037$  through the observations of \hicm\ emission line with the Karl G. Jansky Very Large Array (JVLA) in B-configuration. 
This is  the  second nearest GRB associated with a supernova, and is located  in the outskirts of its large, spiral  host galaxy, identified as 2MASX J11093966-1235116 (LCRS B110709.2-121854) in optical surveys \citep[][]{Shectman96-1996ApJ...470..172S, 2017GCN.22178....1I, Wang18-2018ApJ...867..147W}. 
With  a stellar mass of $\rm 10^{10.1\pm 0.1}\,M_{\odot}$, a star formation rate of   $\rm \sim 1-3\, M_{\odot}\,yr^{-1}$, and a solar metallicity \citep[][]{Perley17-2017GCN.22194....1P, Wang18-2018ApJ...867..147W}, the host galaxy  is amongst the main sequence galaxies  and follows the mass-metallicity relation of  star-forming galaxies in the local Universe \citep[][]{Tremonti04-2004ApJ...613..898T, Brinchmann04-2004MNRAS.351.1151B}. It therefore seems to be a typical star-forming galaxy, hosting a GRB in the nearby Universe, thus an enigmatic case to study.

We describe  the details of observations and data reduction   in  Section \ref{sec:obs}. The results and a detailed discussion  are presented in Section \ref{sec:res}. We summarise our findings  in Section \ref{sec:sum}. 
Throughout this paper we use a flat $\Lambda$CDM  with  $\rm H_0=70\, k\,s^{-1}\, Mpc^{-1}$ and $\rm \Omega_m=0.3$.


\section{Observations and data analysis}
\label{sec:obs}

We used the L-band receivers of the JVLA in B-configuration  to map the \hicm\ emission from the host galaxy of GRB 171205A. 
The observations were carried out on 09-March-2019 and 05-May-2019 for a total time of $\sim$ 10.5 hours (proposal ID: VLA/19A-394; PI: Arabsalmani). 
The observations used the JVLA Software Backend with 16 MHz bandwidth, centred on $\sim$ 1.368 GHz, sub-divided into 4096 channels, yielding a velocity resolution of $\sim$ 0.9 km~s$^{-1}$ and a total velocity coverage of $\sim 3500$~km~s$^{-1}$. 
Two orthogonal linear polarizations were observed, data from which were combined for obtaining the final result.
The bright calibrator 3C286 was observed at the start of each observing run, to calibrate the flux and the system bandpass. 
The secondary calibrator J1130-1449 was also observed intermittently in order to calibrate the time dependent part of the gain and the system bandpass.

``Classic'' {\sc aips} was used for the analysis of the data \citep[][]{Greisen03-2003ASSL..285..109G}. 
The first step was a flagging and calibration loop during which bad visibilities were identified and flagged separately for each of the linear polarizations, following which the antenna-based complex gains were calibrated.
Thereafter system bandpasses were estimated and calibrated for each day's data set separately.
After this, the calibrated data from both the days were combined together.

From the combined data set initially a `channel-averaged' visibility data set was created by averaging together line-free channels,  
which was used  for a standard continuum imaging and self-calibration loop.
3-D imaging was performed on the channel-averaged visibilities with the task {\sc imagr} using 2-D facets, completely covering a circular area of diameter 1.1$^{\circ}$ around the phase centre. 
Imaging was also performed over 0.7 MHz wide channels in order to avoid bandwidth smearing, with the final image produced by averaging the channel-based images.
The total continuum flux was measured to be $\sim$193 mJy. 
We do not detect any continuum emission from the GRB host galaxy, but we detect the continuum emission from  SN 2017iuk / GRB 171205  and measure its flux density to be 2.99 $\pm$ 0.12 mJy using 2-D Gaussian fitting \citep[consistent with the measurements presented in][]{Leung21-2021MNRAS.503.1847L}.  
At the end of the continuum imaging and self-calibration loop, the final antenna-based gains were applied to all the visibilities of the original multi-channel combined data set.

The radio continuum image made using the line-free channels at the end of the self-calibration cycle was used to model and subtract the continuum from the calibrated visibilities in the original multi-channel dataset, using the task {\sc uvsub}.
Any residual spectral baseline across the observed band was then removed using the task {\sc uvlin}. 
We used these residual visibilities after continuum subtraction to create a spectral cube  using the task {\sc imagr}, where the imaging was restricted to the central quarter of the JVLA primary beam. 
The velocity resolution of the cube was optimized to be $\sim 34$ \kms\,  to improve the statistical significance of the detected H{\sc i} 21\,cm emission in independent velocity channels while still having sufficient resolution to accurately trace the velocity field. 
In order to optimize between the signal-to-noise ratio of the detection and the spatial resolution, a robust factor of 0.5 was used to create the cube. 
The synthesised beam for the cube has a Full-Width-at-Half-Maximum (FWHM) of 7.1\arcs$\times$6.3\arcs. 
We reach the theoretical rms-noise of $\sim$ 0.2 mJy per beam in each 34 \kms\, channel. 

We apply the task {\sc momnt} to the spectral cube in order to obtain maps of the H{\sc i} total intensity and the intensity-weighted velocity field for each detected source. 
{\sc momnt} works by masking out pixels in the spectral data cube which lie below a threshold flux in a secondary data cube which the task creates internally within {\sc aips} by smoothing the original cube both spatially and along the velocity axis -- the smoothing ensures that any localized noise peaks are ignored and only emission correlated spatially and in velocity is chosen.
{\sc momnt} created the secondary data cube by applying Hanning smoothing across  blocks of three consecutive velocity channels, whereas spatially a Gaussian kernel of FWHM equal to twelve pixels (about twice the size of the synthesised beam) was applied.
The threshold flux used to select pixels was approximately 1.3 times the noise in a line-free channel of the original cube, a threshold at which noise peaks just start to show up in the total intensity map.
Using  the total-intensity map of \hi\ we  identified the region of emission and obtain the \hi\ spectrum  from the  cube.  
We measured the integrated flux density of \hicm\  emission line  by integrating over the adjacent channels with fluxes above the rms noise, and converted it to the \hi\ mass.

We also use the r-band image of the field of the GRB host galaxy  from the Pan-STARRS1 data archive.


\section{Results and Discussion}
\label{sec:res}

We detect the \hicm\ emission line from the host galaxy of GRB 171205A  at 10$\sigma$ significance, centred on a redshift of $z=0.0371 \pm 0.0001$. We measure  an integrated flux density ($S\Delta v$) of $0.486 \pm 0.048$ Jy \kms\,  and obtain  an \hi\ mass of $\rm 10^{9.49\pm 0.04} M_{\odot}$ for the host galaxy. With a stellar mass of $\rm 10^{10.1\pm0.1}\,M_{\odot}$, the host galaxy has a $\rm M_{HI}/M_*$ of 0.2, the average value for  nearby star forming galaxies with similar stellar masses \citep{Catenilla18-2018MNRAS.476..875C}. Moreover, the \hicm\ line-width   of $\sim$ 300 \kms, measured at the 50$\%$ level of the peak flux,  places the GRB host galaxy on the Tully-Fisher relation in the local Universe \citep[][]{McGaugh12-2012AJ....143...40M}.

\begin{figure*}
\centering
\includegraphics[width=1.0 \textwidth]{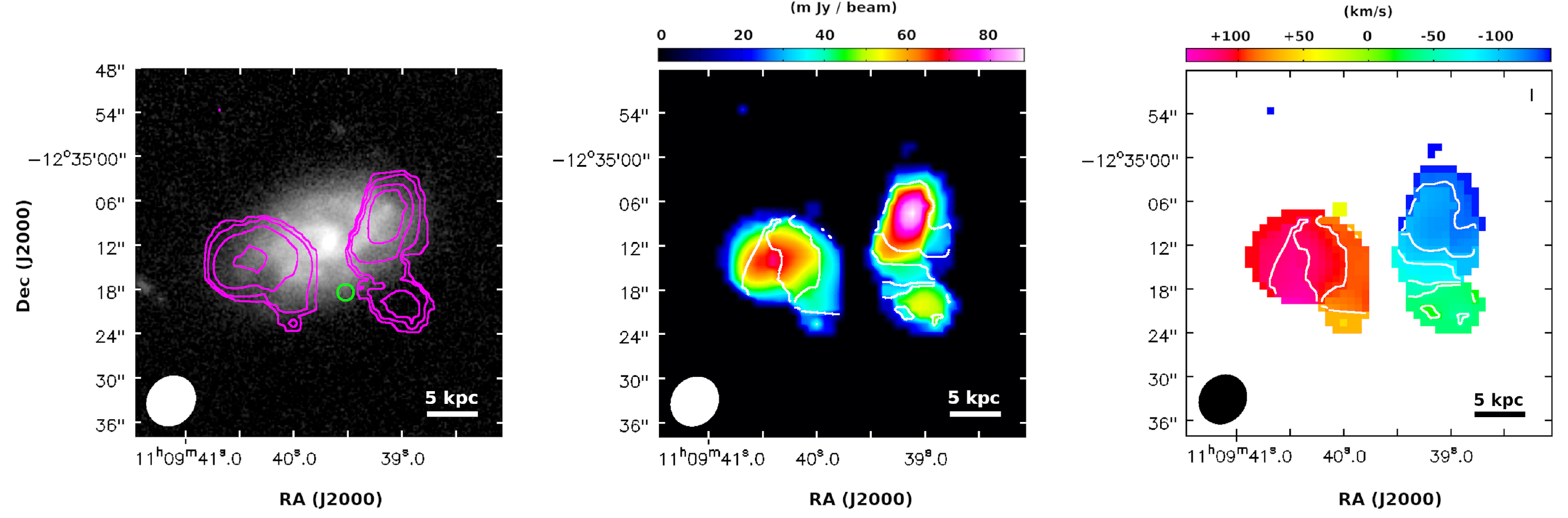}
\caption{ 
{\bf Left panel:} Contours of \hicm\ emission line overlaid on the R-band Pan-STARRS image of GRB 171205A host galaxy. The outermost contour marks the  3-$\sigma$ level of emission in a single channel (with a velocity width of 34 \kms). This is  equivalent to a \hi\ column density of $\rm 5 \times 10^{20}\,cm^{-2}$. Each subsequent contour is  in multiples of $\sqrt{2}$. The green circle marks the position of GRB 171205A. 
{\bf Middle panel:}  Total intensity map of \hicm\ emission line  in color, overlaid with contours of the velocity map of \hicm\ emission line for GRB host. The velocity contours in the left butterfly wing mark the velocities of $+115$, $+95$, and $+75$ \kms\, from left to bottom-right. In the right butterfly wing, the contours mark the velocities of $-25$, $-45$, $-65$, $-85$, $-105$, and $-125$ \kms\, from bottom-left to top. 
{\bf Right panel:}  The intensity-weighted velocity map of \hicm\ emission line  in color, overlaid with velocity contours as in the middle panel. 
The size of the images in all panels  are 50 \arcs $\times$ 50\arcs. The synthesised beam of JVLA observations is shown in the bottom-left corners of the maps.  
\label{fig:host}}
\end{figure*}

Figure \ref{fig:host} shows  the intensity and velocity maps of \hicm\ emission in the host galaxy of GRB 171205A. Given that the  detection limit in  our observations corresponds to   a column density of $\rm 5 \times 10^{20}\,cm^{-2}$ at 3-$\sigma$ significance, we expect the detected \hi\  to be  in the `cold neutral phase' with temperatures of a few hundred Kelvins  \citep[][]{Kanekar11-2011ApJ...737L..33K}. 
The distribution of this cold \hi\  in the host galaxy has a butterfly-shape. The gas associated with the optical disk forms the Northern parts of the wings. 
The Southern parts  consist of extensions of gas outwards of the optical disk of the host where the main disk has no detectable stellar extension. These extensions contribute to more than $20\%$ of the total \hi\ mass in  the galaxy.

Cold \hi\ is absent  from an extended North-South region passing by the optical centre  of the galaxy.  
The  depression of \hi\ in the centre of galaxies  has been typically  observed in nearby spirals and is generally thought to result from the conversion 
of atomic gas to molecular gas in the central regions \citep[][]{Walter08-2008AJ....136.2563W}. However, the depression  is quite unusual in the host galaxy of GRB 171205A: it  is not limited to only the centre of the galaxy,  and it is also  spread over almost half of the velocity width of the \hi.  Deeper observations are planned to confirm the absence of \hi\ at lower column densities  in this region. 
Unlike the case of GRB 980425 and AT2018cow \citep[][]{Arabsalmani15-2015MNRAS.454L..51A, Arabsalmani19-2019MNRAS.485.5411A, Roychowdhury19-2019MNRAS.485L..93R}, GRB 171205A is not located within  the high column density \hi\ ($\rm > 5 \times 10^{20}\,cm^{-2}$) in its host galaxy (see the left panel of  Figure \ref{fig:host} where the GRB position is marked with a green circle). It is instead located close to the east of the western butterfly wing, a side in which  the concentration of \hi\ contours have a sharp edge, suggesting a shock in  the cold \hi.

The   velocity field of \hi\ in the butterfly-shaped distribution  does not show the classical pattern of rotation. 
Gas in the eastern wing is moving away from us, with its velocity in the rest frame of the galaxy  decreasing  from left to the bottom right. Gas in the western wing is moving toward us, with  its velocity in the rest frame of the galaxy increasing  from bottom-left to top. Had the gas been in a rotating disk, we would have seen a continuous change of velocity along an east-west axis going through the centre of the galaxy.

\begin{figure*}
\centering
\includegraphics[width=1.0 \textwidth]{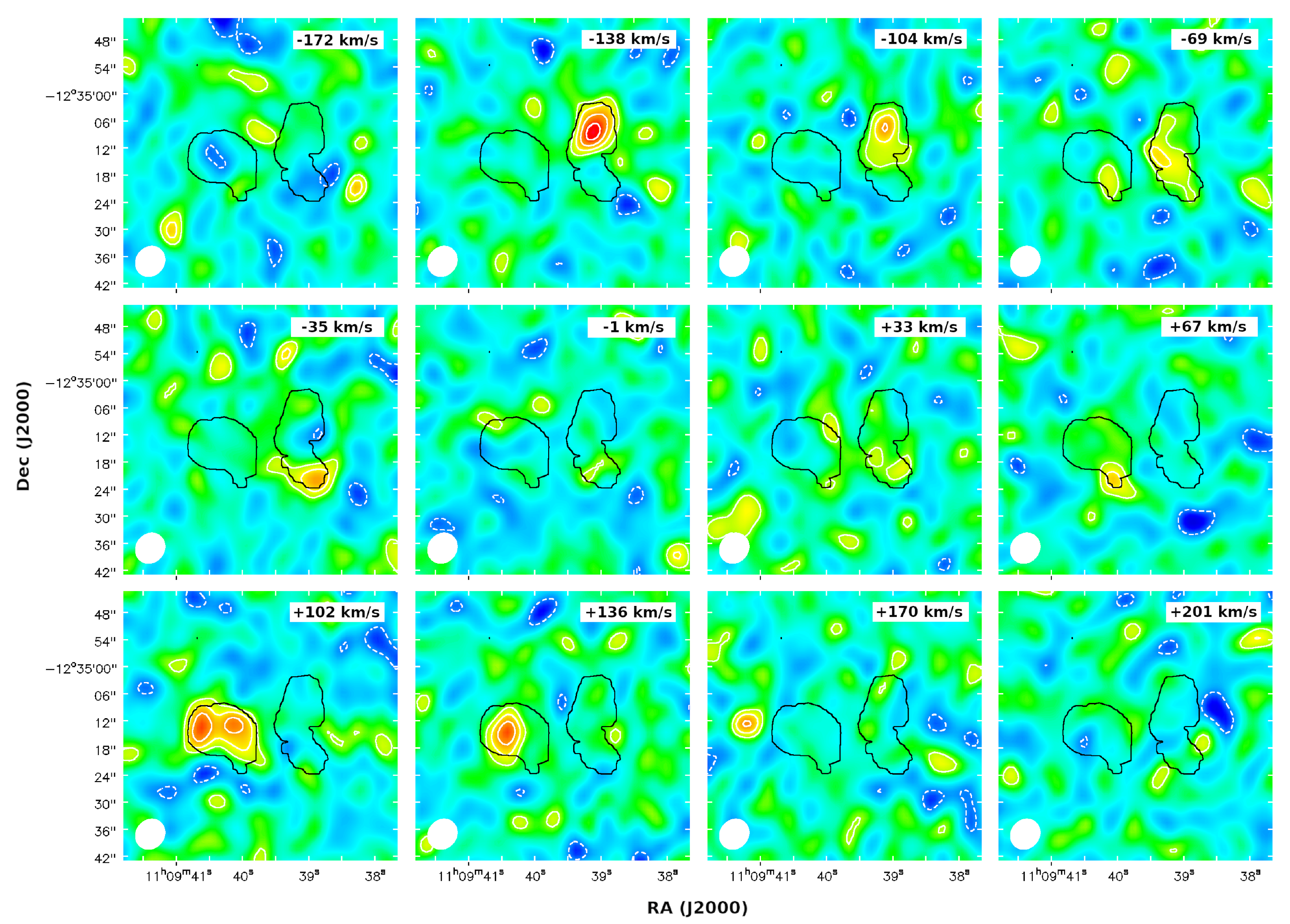}
\caption{ 
The intensity map of the \hicm\ emission in 12  successive channels. Each map has a size of   1 arcmin $\times$ 1 arcmin and a  velocity width of 34 \kms. The contours of the \hicm\ intensity are  overlaid in white with the  dashed contours  at $-2\sigma$ level. The first solid contours are at $2\sigma$ level, with each subsequent contours increasing by $1\sigma$. The black contour marks the total \hicm\ emission (moment-0) from the host galaxy  at 3-$\sigma$ significance. The synthesised beam of JVLA observations is shown in the bottom-left corner of each map. The number in the top-left corner of each map is the central velocity of the channel corresponding to the map, relative to the redshift of the galaxy obtained from \hi.  
\label{fig:chan-map}}
\end{figure*}

This can be better seen in the individual channel maps in Figure \ref{fig:chan-map}. 
The  gas     associated with the optical disk of the galaxy (the Northern parts of the wings) corresponds to a rotating disk, though with missing gas over a large velocity range. This component  is clearly detected in four channels with central velocities of -138, -104, +102, and +136 \kms, but  is not present  in four  channels with central velocities of -35, -1, +33, and +67 \kms. The missing gas  therefore covers  about 150 \kms\ in velocity space. This is an unusually large velocity window  considering that the whole velocity spread of this galaxy is about 300 \kms \citep[see for e.g.,][]{Walter08-2008AJ....136.2563W}.

In order to investigate whether the high column density sensitivity limit of our observations can explain the observed `piling up' of \hi\ in a narrow range of velocities at the edges in the channel maps, we created  models of a rotating \hi\ disk with a flat rotation curve at the edges. Using  the {\sc GALMOD}  task in Gipsy64 \citep[][]{vanderHulst92-1992ASPC...25..131V} we simulated a rotating  \hi\ disk of uniform density and velocity dispersion with large flattened part of the rotation curve at the edges of the disk. We found that irrespective of the inclination angle, even with a large amount of gas  in the outskirts in the flat rotation curve regime,  high column density  \hi\ should be visible  in the central channels owing to projection effects, unlike what we see in our observed channel map. We also simulated a ring of \hi\ gas with no \hi\ present in the inner 2/3rds and with the outer part having almost a flat rotation curve, again with uniform column desnity and velocity dispersion. Even for such an extreme case, the simulated models showed the presence of high column denisty \hi\ in the central channels. We therefore confirm that the apparent deficit of   \hi, associated with the optical disk of the galaxy, in the central channels in Figure \ref{fig:chan-map} is  unlikely to be a result of  the sensitivity limit of the observations.

The extended gas (the southern part of the wings) is present in channels with central velocities between $-35$ and $+33$ \kms\ in Figure \ref{fig:chan-map}, where \hi\ associated with the optical disk of the galaxy  does not show any emission. 
This makes  it  difficult to justify that this gas is merely an extension of the gas disk associated with the optical disk of the galaxy. 
In addition, detection of this component at high column densities ($\rm 5 \times 10^{20}\,cm^{-2}$ at 3$\sigma$  significance), spatially extended over a large area,  is  unlike what is expected for infalling/accreting gas.
It is more  likely  that   an internal/external mechanism has  removed the cold \hi\ from the  central region of the host  and instead has pushed and compressed   it towards the south-western and south-eastern sides.  This hypothesis   is  supported by the high column density  of the extended gas in the southern parts, and also by the pattern of its velocity being similar to that of the gas associated with the optical disk. 
Note that the emission detected in  channels with central velocities of $-69$ and $+102$ \kms\ suggests that the gas in the northern and southern parts of either butterfly wing are dynamically connected,  even when considering the size of the  synthesised beam of the observations.

Feedback from a nuclear  starburst in the host galaxy   can clear  the gas in the central regions and redistribute it  towards the southern/northern part of the galaxy. 
However, the distribution of the  \ha\ emission in the host galaxy  suggests the absence of a nuclear  starburst  (based on the VLT/MUSE observations of the host; Thone et al. in prep.). 
An active galactic nucleus (AGN) could have a similar effect, though over smaller  scales. Nevertheless, the analysis presented in \citet[][]{Wang18-2018ApJ...867..147W} rules out the presence of an AGN in the GRB host. 
Other hydro-dynamical processes such as ram pressure stripping do not explain the \hi\ structure in the galaxy. 
Ram pressure from a dense intergalactic medium  typically pushes large parts of the gas disk towards one side of the galaxy relative to the optical disk, determined by the direction of infall of the galaxy towards the local over density. Such an effect cannot selectively remove gas from the centre of the galaxy. 
In any case we note that the host galaxy is not a part of any identifiable group or cluster based on the optical image of the field, which makes the gas in the galaxy being ram pressure stripped by a dense intergalactic medium an unlikely scenario.

With all above mentioned candidates ruled out,  interaction with a companion remains   the most viable explanation for the peculiar structure of \hi\ in the host galaxy of GRB 171205A. 
While classical tidal interactions  are  expected to result in  gas inflows towards the galaxy centre \citep[e.g.][]{Renaud14-2014MNRAS.442L..33R}, the penetrating passage of a companion  through the disk of the GRB host could  drag the gas out of the galaxy and result in the observed features in the \hi\ distribution. 
Such a passage  creates a tidal perturbation and leads to  large scale shocks  and ram pressure between the gaseous media of the two galaxies. 
The stellar component  is not  directly affected by hydrodynamical processes (shocks, ram pressure). But it should  respond to the change of gravitational potential induced by the displacement of the gas (in this case more than 20$\%$ of the gas budget of the GRB host). 
Furthermore, the tidal perturbation in gas should also  induce  disturbances in the stellar component  (e.g. a warp, tidal tails).  
The  well-ordered distribution of  the stellar component in the GRB host galaxy (see the left panel of Figure \ref{fig:host})  therefore suggests that    the passage is very recent such that the distribution of stars have not \emph{yet} been affected.  

In the  first passage of a low-mass companion,  the gas component of the main galaxy, as a continuous medium, immediately  reacts to the presence of the companion. However, the stellar component which usually has a higher velocity dispersion \citep[][]{Renaud21-2021arXiv210600020R} takes slightly longer to shape an  organized and coherent structure like tidal tails. A few Myr after the pericentre passage, it could well be that  both components experience the tidal effect of the companion, but that only the gas shows a different morphology, while the stars are still being accelerated and are soon to form tidal structures. Such a delay has been seen in  simulations of galaxy interactions \citep[e.g.][]{Renaud21-2021MNRAS.503.5846R}. 
In addition to the mentioned delay, the mass of the companion can play a role: too massive a companion would have affected the stars earlier during its approach, while too light a one would not have created the observed significant
disturbance in the gas. 
An exploration of the parameter space with simulations is required to reach a precise estimate which is out of the scope of this paper, but 
minor mergers can create the  perturbations we seek \citep[][]{DiMatteo07-2007A&A...468...61D, Renaud09-2009ApJ...706...67R}. 
Follow-up simulation studies and observations of this system are planned in order to investigate the  viability of the passage scenario. 

The closest galaxy to the GRB host and thus a possible culprit in the 
passage scenario,  as  identified in optical surveys  is LEDA 951348. This galaxy, with a redshift of $z = 0.0369$ obtained from our \hicm\ observations, is located   at a projected distance of $\sim$ 190 kpc to the North-West of the GRB host, too far for a recent encounter with the GRB host.  
We also search for the presence of   \hi\ knots associated with possible companions with no or faint stellar components in the vicinity of the GRB host galaxy. We tentatively detect two \hi\ knots in the western and eastern sides of the GRB host, both at  projected distances of about 30 kpc from the centre of the host galaxy, but disregard them in our analysis given the low significance of their detection. 
The  closest \hi\ source to the GRB host detected with enough significance  is at a  projected distance of $\sim$ 42 kpc from the GRB host centre, to its North-West, and with  no identified optical counterpart. 
The \hi\  mass of this source ($\rm 10^{8.38\pm 0.08} M_{\odot}$) 
is sufficient to cause  the observed disturbance in the gas disk of the GRB host without yet affecting its stars, but    
the   distance between the \hi\ knot and  the GRB   host is again too large for a recent encounter and would be consistent with timescales $\gtrsim$ 100 Myr. 
Even  a rapid encounter (with a velocity of $\sim$ 500 \kms),  which might explain the absence of tidally disturbed gas in the GRB host, needs to have occurred at least  80 Myr ago 
to explain the $\sim$ 40 kpc distance of the knot.

It is possible that the \hi\  clump in the far South-West of the  butterfly wing   is the remnant of a companion and the gas that has been displaced from the centre of the disk. This clump has a  mass of  $\rm 10^{8.66\pm 0.07} M_{\odot}$, about 15$\%$ of the \hi\ mass and 3$\%$  of the baryonic mass in the GRB host.  
With a trajectory roughly running from North to South, and approximately in the plane of the sky, the companion would  remove the gas in the central regions and create a shock in the main galaxy. This would  explain the sharp edge of the \hi\ distribution on the eastern side of the western wing. 
It  would also cause  the discontinuity  in the   \hi\ velocity field. 
Though the measured line-of-sight velocity of the gas in the clump is close to zero, it can have a large velocity component in the plane of the sky.
Keeping in mind the uncertainties on the inclination of the disk, and of the orbit,  an encounter speed of $\sim$ 500   \kms\, would be compatible with a pericentre passage  less than 20  Myr ago. 
In classical tidal interactions, the SFR is only starting to rise this early after the pericentre passage, because most of the gas is still being compressed and  the collapse of the gas clouds has just started \citep[e.g.][]{Renaud14-2014MNRAS.442L..33R}. 
But in a collision, 
the created shock  can lead to a much more  violent formation of stars.  
Such extreme physical conditions are compatible with the formation of very massive stars. 

GRBs are typically  found in the bright H{\sc ii} regions within the central  $\sim$kpc of their host galaxies \citep[e.g.][]{Fruchter06-2006Natur.441..463F, Lyman17-2017MNRAS.467.1795L}. 
The location of GRB 171205A in the outskirts of its  host is therefore quite atypical.     
This could be explained by the formation of the  massive star progenitor(s) of GRB 171205A in extreme conditions created by a violent shock, in the the vicinity of the gas clump. Note that  the location of GRB 171205A  is at the edge of what would be a shock, as indicated by the structure of \hi\ in the galaxy  (see Figure \ref{fig:host} where the GRB position  is marked with a green circle).

\section{Summary}
\label{sec:sum}

We present a detailed study of the distribution and kinematics of atomic hydrogen in the host galaxy of GRB 171205A through the \hicm\ emission line observation with the JVLA. 
While the global properties of stars and gas in the host galaxy  of GRB 171205A  appear quite normal,  the structure of its \hi\ shows very unusual features. \hi\ is  absent  from an extended North-South region passing by the optical center of the galaxy, but  extends   outwards  the optical disk in  the south in both sides, with a  clear discontinuity in its velocity field  along the major axis of the galaxy.   We rule out internal (hydrodynamical) effects as the cause of these peculiarities  given the absence of enhanced star formation or an AGN in the centre of the galaxy. The absence of gas in the central region of the galaxy and the well-ordered stellar distribution in the GRB host rules out past tidal interactions as the cause. 
The most viable remaining explanation  is a very recent (only a few Myr ago) collisional passage of a companion through the disk of the GRB host such that the distribution of stars have not \emph{yet} been affected. 
It is possible  that the \hi\  clump that we detect in the far South-West of the  gas distribution of the GRB host  is the remnant of the companion and the  gas that has been displaced from the centre of the disk. 
The location of GRB 171205A  is consistent with the formation of its progenitor star(s) due the shock induced by the collision that is responsible for the observed peculiar features in the \hi, supporting the idea  that these rare explosions could be ignited by rare dynamics that result in extreme conditions. 
Much more detailed studies must be carried out to validate  this idea.


\section*{Acknowledgments}
M.A. thanks   Lister Stavely-Smith  for helpful discussions.  
The presented study is funded by the Deutsche Forschungsgemeinschaft (DFG, German Research Foundation) under Germany´s Excellence Strategy – EXC-2094 – 390783311. 
S.R. acknowledges support from the ESO Scientific Visitor Programme. 
F.R. acknowledges support from the Knut and Alice Wallenberg Foundation. 
The National Radio Astronomy Observatory is a facility of the National Science Foundation operated under cooperative agreement by Associated Universities, Inc. 


\end{document}